# Variation of singing styles within a particular *Gharana* of Hindustani classical music – a nonlinear multifractal study

**Archi Banerjee[1,2]*, Shankha Sanyal[1,3], Ranjan Sengupta[1] and Dipak Ghosh[1]**

[1]*Sir C.V. Raman Centre for Physics and Music,*
*Jadavpur University, Kolkata - 700032*
[2] *Rekhi Centre of Excellence for the Science of Happiness,*
*Indian Institute of Technology Kharagpur, Kharagpur - 721302*
[3] *School of Languages and Linguistics,*
*Jadavpur University, Kolkata - 700032*
**corresponding author, email: archibanerjee7@gmail.com*

## ABSTRACT

Hindustani classical music is entirely based on the "*Raga*" structures. In Hindustani music, a "*Gharana*" or school refers to the adherence of a group of musicians to a particular musical style. *Gharanas* have their basis in the traditional mode of musical training and education and every *Gharana* has its own distinct features; though within a particular *Gharana*, significant differences in singing styles are observed between generations of performers, which can be ascribed to the individual creativity of that singer. This work aims to study the evolution of singing style among four artists of four consecutive generations from *Patiala Gharana*. For this, *alap* and *bandish* parts of two different *ragas* sung by the four artists were analyzed with the help of non linear multifractal analysis (MFDFA) technique. The multifractal spectral width obtained from the MFDFA method gives an estimate of the complexity of the signal. The observations from the variation of spectral width give a cue towards the scientific recognition of *Guru-Shisya Parampara* (teacher-student tradition) – a hitherto much-heard philosophical term. From a quantitative approach this study succeeds in analyzing the evolution of singing styles within a particular *Gharana* over generations of artists as well as the effect of globalization in the field of classical music.

## 1. INTRODUCTION

### 1.1. A brief introduction to Hindustani classical music

Indian classical music can be compared to the endless sky when its creative aspect is considered. It is believed that in ancient times, there existed a single form of the style of Indian Classical Music. However, the consecutive foreign invasions over the centuries had a great impact on the Indian Classical Music and this created a division into this form of music. This lead to the regeneration of two forms of Indian Classical Music: the Carnatic Music (Usually this style is followed in the Southern parts of India and considered by some as the basic version of Indian Classical Music) and the Hindustani Music (The North Indian and the improvised version of the Indian Classical Music). *Raga* is the heart of Hindustani classical music. Each *Raga* has a well defined structure consisting of a series of four/five or more musical notes upon which its melody is constructed. However, the way the notes are approached and rendered in musical phrases and the mood they convey are more important in defining a *Raga* than the notes themselves [1]. This leaves ample scope for improvisation within the structured framework of any *Raga*. Every performer of this genre (Hindustani classical music) is essentially a composer as well as an artist. While performing a *Raga* every time, an artist gradually goes on to explore new pathways – new connections between the used notes following his current state of imagination but, at the same time, he maintains the basic discipline of that *Raga*. In this way he slowly moves beyond the strangle hold of the *Raga* grammars and dives into the more subtle and sublime emotional self of the *Raga.*

*Alap* is the opening section of a *Raga* performance in typical Hindustani classical (*Khayal*) singing/instrument playing style and acts as the preface of the *Raga*. In the *alap* part the *Raga* is introduced and the paths of its development are revealed using all the notes used in that particular *Raga* and allowed transitions between them with proper distribution over time. *Alap* is usually accompanied by the tanpura drone only and sung at a slow tempo or sometimes without tempo. Then, in case of vocal *Raga (Khayal)* performances, comes the *vilambit bandish* part where the lyrics and rhythmic cycle or *taal* are introduced. *Bandish* is a fixed, melodic composition in Hindustani vocal music, set in a specific *raga*, performed with rhythmic accompaniment by a tabla or pakhawaj, a steady drone, and melodic accompaniment by a sarangi, harmonium etc. *Vilambit* is a type of *bandish* which is sung at a very slow tempo, or *laya*, of 10-40 beats per minute. The first paragraph of the song – *Sthayi* is followed by the second one – *Antara* [2]. A *Vilambit bandish* is usually followed by a *Drut* (or higher tempo ~90-120 bpm) *bandish* along with several types of melodic and lyrical improvisations within both the *Vilambit* and *Drut bandishes.*

### 1.2. *Gharana* tradition of Hindustani classical music and *Guru-Shishya parampara*

Beyond this infinite canvas of individual improvisations, some different styles of presentation are also observed across the country while rendering any particular *raga*. In the context of Hindustani classical music, the major differences in the *Raga* presentation styles are named after different "*Gharanas*". The coinage "*Gharana*" came from the Hindi word "Ghar" (House). It is commonly observed that the *Gharanas* are named after different places, viz., Agra, Patiyala, Gwalior, Maihar, Bishnupur, Indore etc. The naming of these *Gharanas* mostly indicates towards the origination of these particular musical styles or ideologies. Under the Hindustani Classical Music, the tradition of "*Gharana*" [3, 4] system holds special importance for many listeners. Perhaps, this feature is so unique that no where around the world can one find this sought of a tradition. The *Gharana* system is followed by both the North-Indian as well as the South-Indian forms of Indian classical music. In south India, the term *Gharana* is acknowledged by the word "Sampraya". In ancient times, there existed several Samprayas such as the "Shivmat", the "Bhramamat" and the "Bharatmat" [5]. *Gharanas* have their basis in the traditional mode of musical training and education. Every *Gharana* has its own distinct features; the main area of difference between *Gharanas* being the manner in which the notes are sung. Though, in most cases, even within a particular *Gharana*, significant changes in singing styles are observed between generations of performers. These changes are ascribed to the individual creativity of that singer, which has led to the evolution of the singing style of that particular *Gharana*.

One of the most unique and exclusive feature which is inherent to the teaching process of Indian Classical Music is known as the "*Guru-Shishya*" tradition [6]. History and statistics reveal that majority of the finest artists of Indian Classical Music have been produced through the *Guru-Shishya Parampara* (tradition). In India, the *Gharana* system has contributed to all the three forms of music, that is vocal, instrumental and dance. The *Gharana* comes into existence through the confluence of the "*Guru*" and the "*Shishya*" [7]. A wise "*Guru*" through his intelligence, aptitude and shear practice creates a sense of uniqueness and exclusivity and thereby inculcates a special eminence into his form of music. These attributes and traits are amicably transferred into the talented "*Shishya*" and the particular form of the performing arts thus becomes a tradition. These exceptional qualities are in fact so strong and prominent that the audiences can immediately recognize the *Gharana* i.e the similarity in singing/dancing/playing style between two artists. It is believed that when so ever the form or style created by the founder "*Guru*" is carried forth till three generations; it turns in to the form of "*Gharana*". The name of the *Gharana* can be same as the name of the founder "*Guru*", or came be named after the place where the founder "*Guru*" resided. Throughout the history of the Hindustani classical music tradition, students were often born into a musical family practicing the art of music in the *Guru-Shishya* tradition, which was passed down through hereditary means by the musically gifted members of the family [8]. Traditionally, families belonging to a *Gharana* practiced the art of relaying musical knowledge from one generation to the next, and the music and the particular style of one *Gharana* became the basis for playing, understanding, and critiquing music within the Hindustani classical style. Over time the *Gharana* system has expanded to include a student base that was not necessarily connected through the blood line of the *Guru*. The tradition shifted to accommodate these new students, but only so far as to welcome them into the system that was already in place: the student, though unrelated by blood, still became a part of the *Guru*'s family. This practice still remains in certain areas of India, depending on the *Guru*, but with the constant influence of global ideology, the "*Guru-Shishya*" system has had to adapt once more. The students may not be able to live with their *Guru* throughout their lives, but the respect, the devotion and the support offered to the *Guru* remains the same.

### 1.3. Earlier studies on *Gharana* tradition and background of the present study

Now-a-days, the "*Gharana*" system in Hindustani classical music is affine to a number of ambiguous ideas among artists. According to a few, *Gharana* system exists pretty distinctively, while some are against this thought. What defines the *Gharana*? Is it really true that the artists of the same *Gharana* keep their unique singing style unchanged over generations or evolution of music takes place like everything else in nature? These questions are still unanswered; at least from a scientific point of view. From the earlier discussions it becomes clear that *Gharana* represents a family of musicians, a well-knit unit evolving, guarding and disseminating the distinctive style through its members, some of whom are well-known performers and some who are not. The wealth of the *Gharana* has always been its knowledge – while performing a *raga* the knowledge of song lyrics, rhythm, ornamentation, which coupled with distinctive voice production, phrasing and sequence, produces a unified whole, an aesthetic form special to each family, a unique lineage of features ascribed to a particular *Gharana.* The literature has very few scientific studies which attempted to analyze the significant musical styles that define a particular *Gharana* [9, 10] or the differences in *Raga* presentation styles observed between different *Gharanas.* Most of the previous studies looked into the aesthetic and prosodic features [11] that distinguish one *Gharana* from another. Datta et.al [11] made use of projection pursuit techniques to analyze the similarity between different artists, meant to classify the ragas objectively rather than perceptually. It took the help of various linear features such as MFCC, RMS energy Spectral Irregularity/ Centroid to identify the specific features in singers of a particular *Gharana.* This domain of research was still lacking a rigorous scientific study which can objectively try to look for the similarities and dissimilarities in the overall singing style of different artists from the same *Gharana.* To begin our search in such a huge field of study our initial endeavor was to perform an analytical comparison between the renditions of two particular *ragas* sung by some artists who are said to belong to a particular *Gharana* of Hindustani classical music. In this study, for the first time, we applied robust non-linear tools to extract multifractal features from the complete audio signal waveforms of the *Raga* renditions sung by four generations of singers belonging to a particular *Gharana (Patiyala).* The multifractal techniques used in our study have been discussed in depth in the next section.

### 1.4. Use of multifractal technique (MFDFA) to identify different singing styles

Previous knowledge suggests that music signals have a complex behavior: at every instant components (in micro and macro scale: pitch, timbre, accent, duration, phrase, melody etc) are closely linked to each other [12, 13]. These properties are peculiar of systems with chaotic, self organized, and generally, nonlinear behavior. Therefore, the analysis of music using linear and deterministic frameworks seems unrealistic and a non-deterministic/chaotic approach is needed in understanding the speech/music signals. Music data is a quantitative record of variations of a particular quality (displacement) over a period of time. One way of analyzing it is to look for the geometric features to help towards categorizing the data in terms of concept [14]. The time evolution of the inherent geometrical structure of a certain music piece can be judged rigorously using latest-state-of-the-art nonlinear technique – fractal analysis, which determines the symmetry scaling behavior of a time series. Fractal analysis of audio signals was first performed by Voss and Clarke [15], who analyzed amplitude spectra of audio signals to find out a characteristic frequency $f_c$, which separates the white noise from a highly correlated behavior ($\sim 1/f^2$). However, it is well-established experience that naturally evolving geometries and phenomena are rarely characterized by a single scaling ratio; different parts of a system may be scaling differently, i.e., the clustering pattern is not uniform over the whole system. Such a system is better characterized as 'multifractal' [16]. A multifractal can be loosely thought of as an interwoven set constructed from sub-sets with different local fractal dimensions. Real world systems are mostly multifractal in nature. Music too, has non-uniform property in its movement [17, 18] as it is often featured by very irregular dynamics, with sudden and intense bursts of high-frequency fluctuations. Su & Wu [17] showed that both melody and rhythm can be considered as multifractal objects by separating both of them as series of geometric points. Live performances encompass a variety of such musical features including tempo fluctuations [19], notation and timbre variation to name a few. To study such a signal, Multifractal Detrended Fluctuation Analysis (MFDFA) would certainly be a better tool than Detrended Fluctuation Analysis (DFA) as DFA measures the monofractal scaling property of a time series and yields only a single scaling exponent. The MFDFA technique gives us the multifractal spectral width (W) which is a measure of the inherent complexity of the music signal. The MFDFA was first conceived by Kantelhardt et.al [20] as a generalization of the standard DFA [21]. We hypothesize that the change of multifractal spectral width of the signal will give us a cue about the variation of singing style among the artists belonging to different *Gharanas* as well as artists from successive generations representing the same *Gharana*. Four popular vocalists of four successive generations from a particular *gharana* (Patiala) of Hindustani music were chosen for this study. The renditions of two well known basic *raga*s – *Bageshri* & *Jaijawanti* (containing both *alap* and *vilambit bandish* parts), sung by those four vocalists were analyzed using MFDFA technique and the spectral width values corresponding to different parts of the *Raga* renditions were compared for the four artists to study the evolution in their singing styles.

## 2. EXPERIMENTAL DETAILS

### 2.1. Choice of Music Signals

In this work, our objective was to study the similarity & changes in the singing pattern of a particular *raga* over generations of artists of the same *Gharana*. For the experiments, recordings of the renditions of two *raga*s (*Bageshri* & *Jaijawanti*), containing the introductory *alap* section as well as both the *sthayi* and *antara* parts of the low tempo *vilambit bandish*, were collected from four artists of consecutive generations belonging to a particular *Gharana* (*Patiyala*) of Hindustani classical music. For each *raga* the chosen *bandish* was same for all the four artists.

### 2.2. Processing of Music Signals

All the signals are digitized at the rate of 44100 samples/sec 16 bit format. The *alap* part, *sthayi* and *antara* of the *bandish* part were cut out separately from each rendition before detailed analysis. It is expected that in the *alap* part, note combinations or improvisations will differ for different vocalists while establishing the *raga* and hence there was significant variation in the length of the *alap* pieces chosen for our analysis. So, to minimize the variation due to improvisations, in case of each vocalist, about 20 seconds of the *alap* part were cut out which led only to identification of the *raga*. The said 20 seconds clips were selected by an eminent musician with more than 20 years of experience in performing Hindustani classical music. On the contrary, the *bandish* part has lesser chances of variation in note combinations as for a particular *raga*, the same *bandish* was sung by all the vocalists keeping the melody structure almost same. Although for different vocalists significant variations in the scansion of the *bandish* i.e. the distribution of the lyrics over the whole rhythm cycle of the *taal* are expected. These clips were then selected for analysis.

## 3. METHODOLOGY

For the analysis each of the chosen *alap* parts of 20 seconds duration was divided into 4 equal parts of 5 seconds and their multifractal spectral widths were calculated using the MFDFA technique. The variation in the spectral widths among the 4 vocalists of 4 different generations was observed separately for the 2 chosen *raga*s. Similar observations were found in case of the *sthayi* and *antara* part of the *vilambit bandish*es. The detailed algorithm for MFDFA technique is given in Section 3.1.

### 3.1. Multifractal Detrended Fluctuation Analysis of sound signals

The time series data obtained from the sound signals are analyzed using MATLAB and for each step an equivalent mathematical representation is given which is taken from the prescription of Kantelhardt et al [20]. In MFDFA technique, first the noise-like structures of the signal was converted into a random walk like signals. It can be represented as:

$$Y(i) = \sum (x_k - \bar{x}) \qquad (1)$$

where $\bar{x}$ is the mean value of the signal.

The integration reduced the level of noise present in experimental records and finite data. Then the whole length of the signal is divided into $N_s$ no of non-overlapping 'intervals' (int) consisting of certain no. of samples. For 's' as sample size and 'N' the total length of the signal, the intervals are:

$$N_s = \text{int}\left(\frac{N}{s}\right) \quad (2)$$

Since the length N of the series is often not a multiple of the considered time scale s, a short part at the end of the profile may remain. In order not to disregard this part of the series, the same procedure is repeated starting from the opposite end. Thereby, $2N_s$ segments are obtained altogether. The local RMS variation for any sample size 's' is the function F(s, ν). This function can be written as follows:

$$F^2(s,\nu) = \frac{1}{s}\sum_{i=1}^{s}\{Y[(\nu-1)s+i] - y_\nu(i)\}^2 \quad (3)$$

For $\nu = N_s + 1........, 2N_s$, where $y_\nu(i)$ is the least square fitted value in the bin ν. In this work, a least square linear fit using first order polynomial (MF-DFA -1) is performed. The q-order overall RMS variation for various scale sizes can be obtained by the use of following equation

$$F_q(s) = \left\{\frac{1}{N_s}\sum_{\nu=1}^{N_s}[F^2(s,\nu)]^{\frac{q}{2}}\right\}^{\left(\frac{1}{q}\right)} \quad (4)$$

where q is an index that can take all possible values except zero, because in that case the factor 1/q is infinite. The scaling behavior of the fluctuation function is obtained by drawing the log-log plot of $F_q(s)$ vs. s for each value of q.

$$F_q(s) \sim s^{h(q)} \quad (5)$$

In general, the exponent h(q) may depend on q. For stationary time series, h(2) is identical to the well-known Hurst exponent 'H'. Thus, we will call the function h(q) generalized Hurst exponent (**Fig. 1a**). The Hurst exponent is measure of self-similarity and correlation properties of time series produced by fractal. The presence or absence of long range correlation can be determined using Hurst exponent. A monofractal time series is characterized by unique h(q) for all values of q. The generalized Hurst exponent h(q) of MF-DFA is related to the classical scaling exponent τ(q) by the relation

$$\tau(q) = qh(q) - 1 \quad (6)$$

A monofractal series with long range correlation is characterized by linearly dependent q order exponent τ(q) with a single Hurst exponent 'H' (h(q) for q=2). Multifractal signal on the other hand, possess multiple Hurst exponent and in this case, τ(q) depends non-linearly on q [22]. Another way to characterize a multifractal series is the singularity spectrum f(α), which is related to τ(q) via a Legendre transform [23, 24] as follows

$$\alpha = \tau'(q) \quad \text{and} \quad f(\alpha) = q\,\alpha(q) - \tau(q). \quad (7)$$

Here, α is the singularity strength or Hölder exponent, while f(α) denotes the dimension of the subset of the series that is characterized by α. Using Eq. (6), α and f(α) can be directly related to h(q)

$$\alpha(q) = h(q) + qh'(q)$$

$$f(\alpha) = q[\alpha(q) - h(q)] + 1 \quad (8)$$

Here, h'(q) denotes the first derivative of h(q) with respect to q. The f(α) vs α plot or the multifractal spectrum (**Fig. 1b)** is capable of providing information about relative importance of various fractal exponents in the series e.g., the width of the spectrum denotes range of exponents. A quantitative characterization of the spectra may be obtained by least square fitting it to a quadratic function [25] around the position of maximum $\alpha_0$,

$$f(\alpha) = A(\alpha - \alpha_0)^2 + B(\alpha - \alpha_0) + C \quad (9)$$

where C is an additive constant $C = f(\alpha_0) = 1$. B indicates the asymmetry of the spectrum. It is zero for a symmetric spectrum. The width of the spectrum can be obtained by extrapolating the fitted curve to zero. Width W is defined as,

$$W = \alpha_1 - \alpha_2, \qquad \text{with } f(\alpha_1) = f(\alpha_2) = 0 \quad (10)$$

The width of the spectrum gives a measure of the multifractality of the spectrum. Greater is the value of the width W greater will be the multifractality of the spectrum. For a monofractal time series, the width will be zero as h(q) is independent of q.

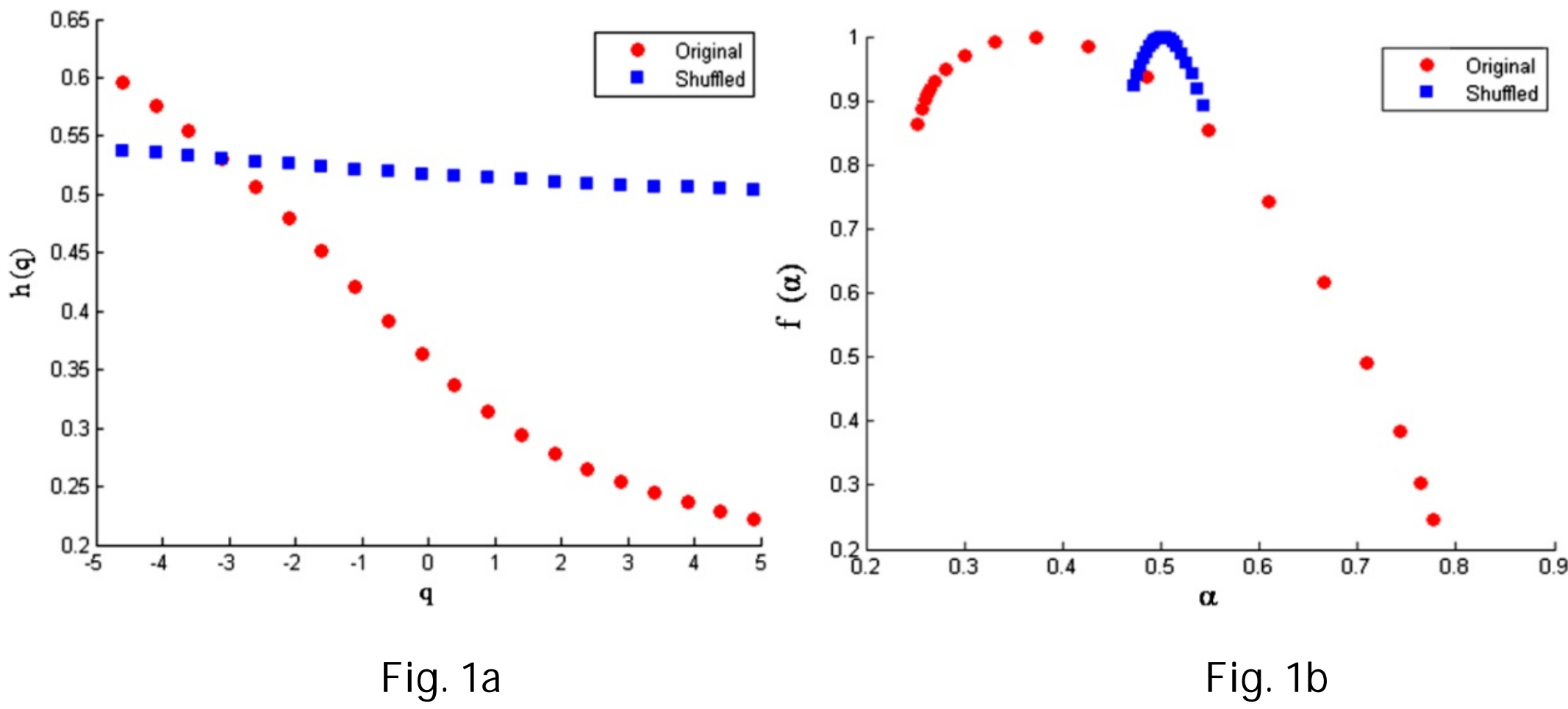

Fig. 1a Fig. 1b

**Fig. 1.** (a) Sample h(q) vs q plot and (b) sample f (α) vs α plot for original and shuffled series corresponding to a specific experimental audio signal.

The origin of multifractality in a sound signal time series can be verified by randomly shuffling the original time series data [26]. In general, two different types of multifractality are present in a time series data: (i) Multifractality due to a broad probability density function for the values of the time series. Here, the multifractality of the time series cannot be removed by random shuffling and the shuffled data has the same variation of h(q) as the original data (ii) Multifractality due to a variety of long-range correlations due to the small and large fluctuations. In this case, the probability density function of the values can be a regular distribution with finite moments, for e. g. a Gaussian distribution. The corresponding shuffled series will exhibit non-multifractal scaling, since all long-range correlations are destroyed by the shuffling procedure. All long range correlations that existed in the original data are removed by this random shuffling and what remains is a totally uncorrelated sequence. Hence, if the multifractality of the original data was due to long range correlation, the shuffled data will show non-fractal scaling. If any series has multifractality both due to long range correlation as well as due to probability density function, then the shuffled series will have smaller width W and hence weaker multifractality than the original time series as is evident from **Fig. 1**.

## 4. RESULTS AND DISCUSSIONS

The multifractal spectral width (W) was computed for each part of the *raga* clips for all the four artists. Higher the value of W, higher is the degree of complexity present in the signal. Thus change in the spectral width due to change of note combinations, transitions between the notes; tempo and rhythm variations in a musical piece would be significantly important to characterize the singing style of different artists. **Table 1** show all the spectral width values where Artist 1 represents the oldest among the 4 artists of *Patiyala Gharana* that we chose to study. Artists 2, 3 and 4 represent singers of consecutive generations from the same *Gharana*.

**Table 1.** Variation of the spectral width values for all 4 artists while rendering 2 chosen *raga*s

| *Raga used* | | | Artist 1 | Shuffled width | Artist 2 | Shuffled width | Artist 3 | Shuffled width | Artist 4 | Shuffled width |
|---|---|---|---|---|---|---|---|---|---|---|
| ***Bageshri*** | ***Alap*** | Part 1 | 0.49 | 0.11 | 0.73 | 0.02 | 0.49 | 0.07 | 0.63 | 0.01 |
| | | Part 2 | 0.63 | 0.09 | 0.72 | 0.03 | 0.42 | 0.03 | 0.63 | 0.03 |
| | | Part 3 | 0.77 | 0.07 | 0.86 | 0.01 | 0.54 | 0.09 | 0.53 | 0.03 |
| | | Part 4 | 0.46 | 0.11 | 0.75 | 0.02 | 0.65 | 0.07 | 0.83 | 0.06 |
| | ***Bandish Sthayi*** | Part 1 | 0.37 | 0.09 | 0.82 | 0.02 | 0.72 | 0.06 | 0.41 | 0.03 |
| | | Part 2 | 0.65 | 0.08 | 0.65 | 0.05 | 0.62 | 0.05 | 0.62 | 0.01 |
| | | Part 3 | 0.45 | 0.08 | 0.77 | 0.07 | 0.59 | 0.11 | 0.41 | 0.03 |
| | | Part 4 | 0.48 | 0.10 | 0.66 | 0.06 | 0.65 | 0.09 | 0.34 | 0.03 |
| | ***Bandish Antara*** | Part 1 | 0.28 | 0.13 | 0.49 | 0.03 | 0.47 | 0.09 | 0.48 | 0.02 |
| | | Part 2 | 0.55 | 0.09 | 0.67 | 0.04 | 0.54 | 0.05 | 0.55 | 0.02 |
| | | Part 3 | 0.59 | 0.09 | 0.66 | 0.04 | 0.56 | 0.10 | 0.46 | 0.03 |
| | | Part 4 | 0.46 | 0.16 | 0.49 | 0.07 | 0.69 | 0.06 | 0.41 | 0.01 |
| ***Jaijawanti*** | ***Alap*** | Part 1 | 0.57 | 0.08 | 0.87 | 0.08 | 0.30 | 0.02 | 0.60 | 0.04 |
| | | Part 2 | 0.91 | 0.25 | 0.84 | 0.04 | 0.63 | 0.01 | 0.64 | 0.02 |
| | | Part 3 | 0.72 | 0.07 | 0.57 | 0.07 | 0.55 | 0.08 | 0.61 | 0.04 |
| | | Part 4 | 0.90 | 0.09 | 0.86 | 0.04 | 0.56 | 0.04 | 0.56 | 0.02 |
| | ***Bandish Sthayi*** | Part 1 | 0.79 | 0.08 | 0.78 | 0.05 | 0.27 | 0.07 | 0.56 | 0.02 |
| | | Part 2 | 0.90 | 0.03 | 0.92 | 0.05 | 0.37 | 0.05 | 0.52 | 0.02 |
| | | Part 3 | 0.79 | 0.07 | 0.62 | 0.03 | 0.36 | 0.05 | 0.37 | 0.01 |
| | | Part 4 | 0.72 | 0.06 | 0.58 | 0.02 | 0.26 | 0.07 | 0.40 | 0.05 |
| | ***Bandish Antara*** | Part 1 | 0.58 | 0.03 | 0.77 | 0.05 | 0.30 | 0.05 | 0.37 | 0.02 |
| | | Part 2 | 0.56 | 0.04 | 0.88 | 0.02 | 0.52 | 0.03 | 0.29 | 0.01 |
| | | Part 3 | 0.61 | 0.05 | 0.50 | 0.03 | 0.15 | 0.08 | 0.41 | 0.04 |
| | | Part 4 | 0.59 | 0.01 | 0.71 | 0.05 | 0.37 | 0.09 | 0.47 | 0.03 |

From **Table 1** it is evident that the average multifractal spectral width is higher in case of *Raga Jaijawanti* than that of *Raga Bageshri* for all 4 artists both in the *alap* part as well as the entire *bandish* part. Thus we can conclude that the overall complexity of *Raga Jaijawanti* is higher than that of *Raga Bageshri*. To make the trends in our data easier to visualize, variation of multifractal spectral widths among the four generations of artists from *Patiyala Gharana* while singing different parts of the renderings of *Raga Bageshri* and *Raga Jaijawanti* were plotted separately in **Fig. 2.** The following figures – **Fig. 2a** and **Fig. 2b** represent the variation of multifractal spectral width (W) for the 4 artists while rendering *alap* parts of *Raga Bageshri* and *Raga Jaijawanti* respectively. **Fig. 2c** and **Fig. 2d** represent the same for the *Bandish Sthayi* parts whereas **Fig. 2e** and **Fig. 2f** represent the same for *Bandish Antara* parts corresponding to the chosen two *Ragas*. The error bars given in all the following figures represent the computational errors introduced in the multifractal algorithm used in this work.

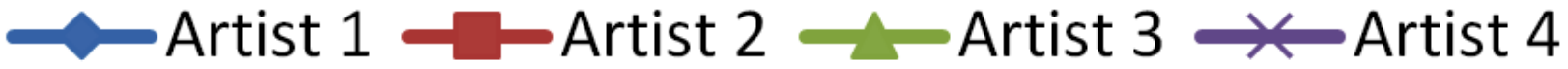

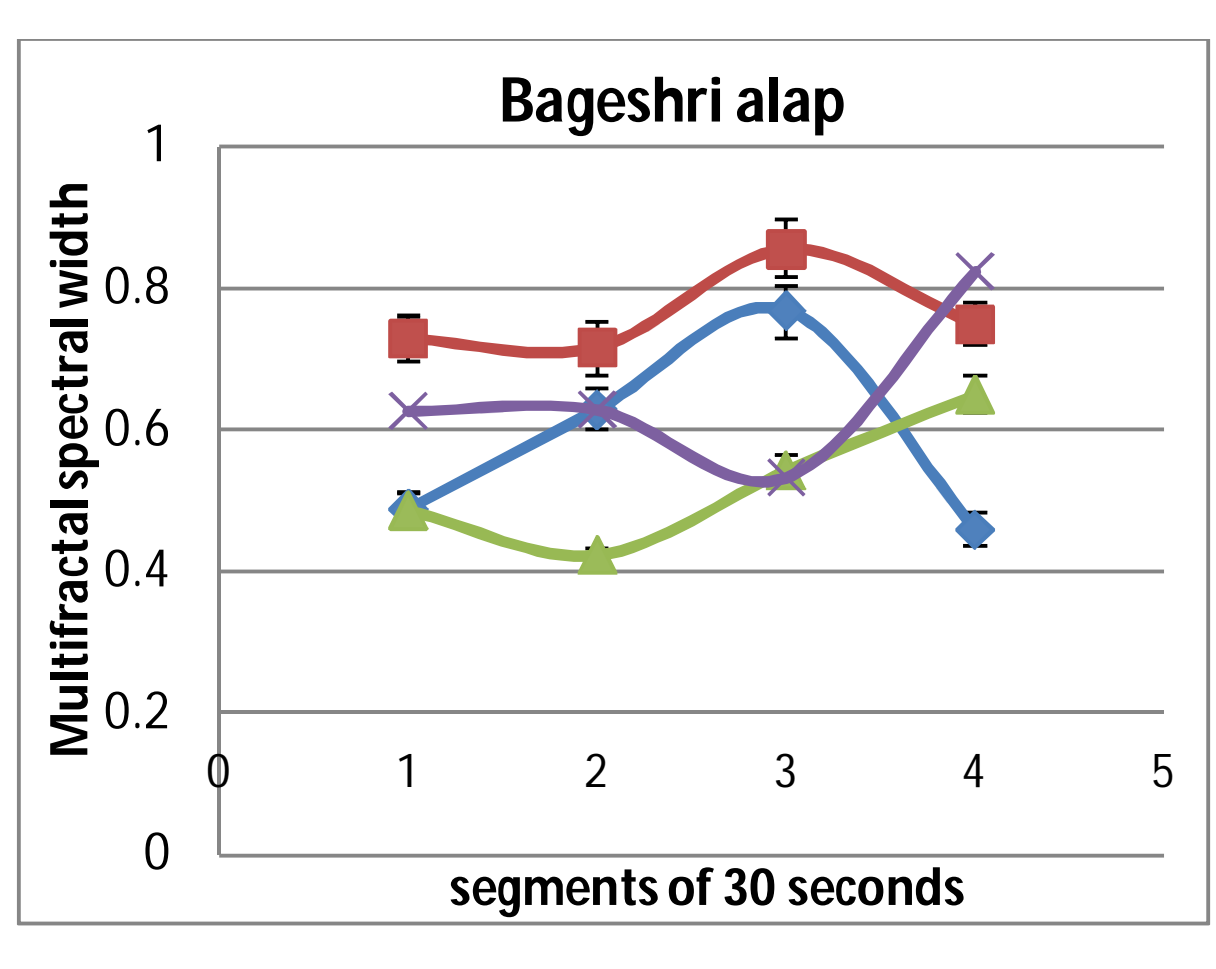

Fig. 2a

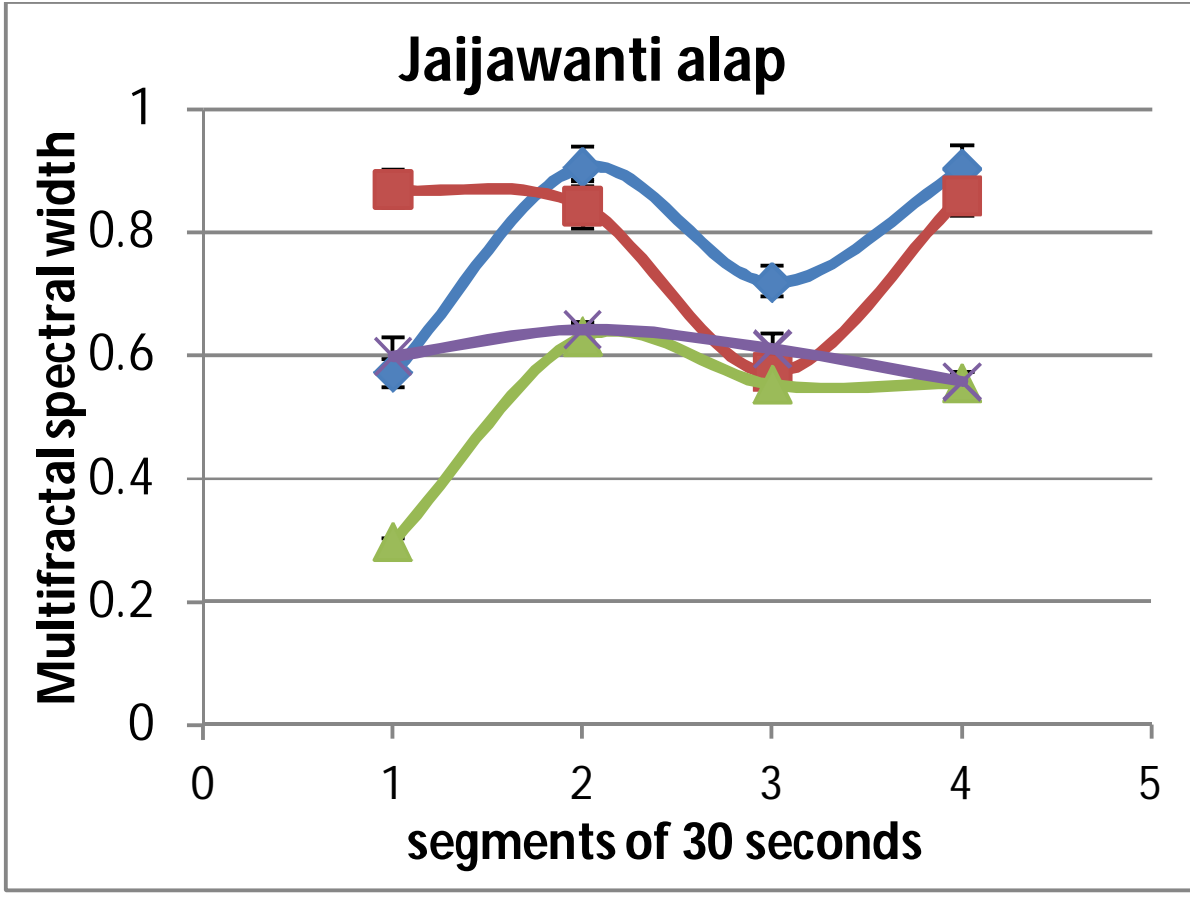

Fig. 2b

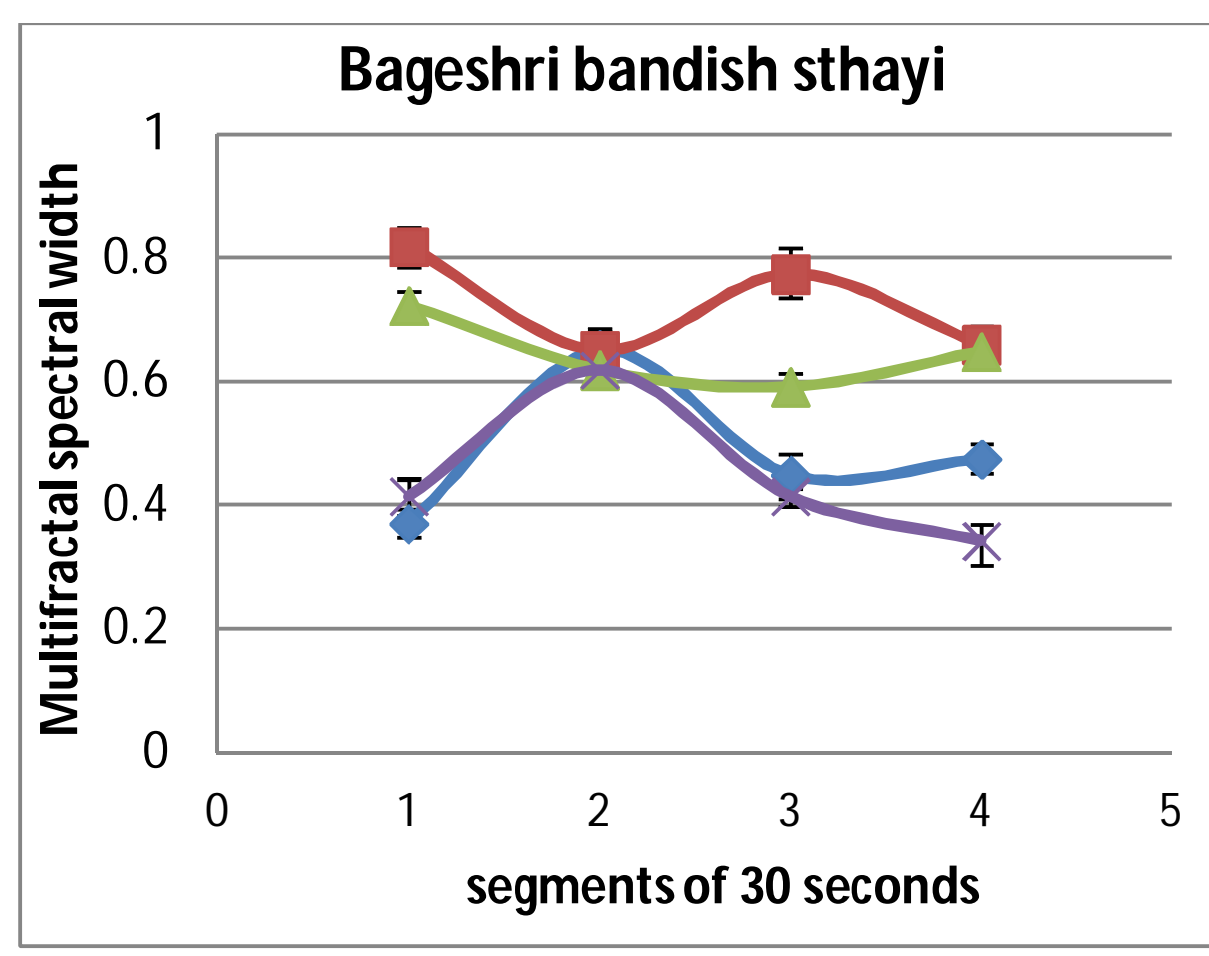

Fig. 2c

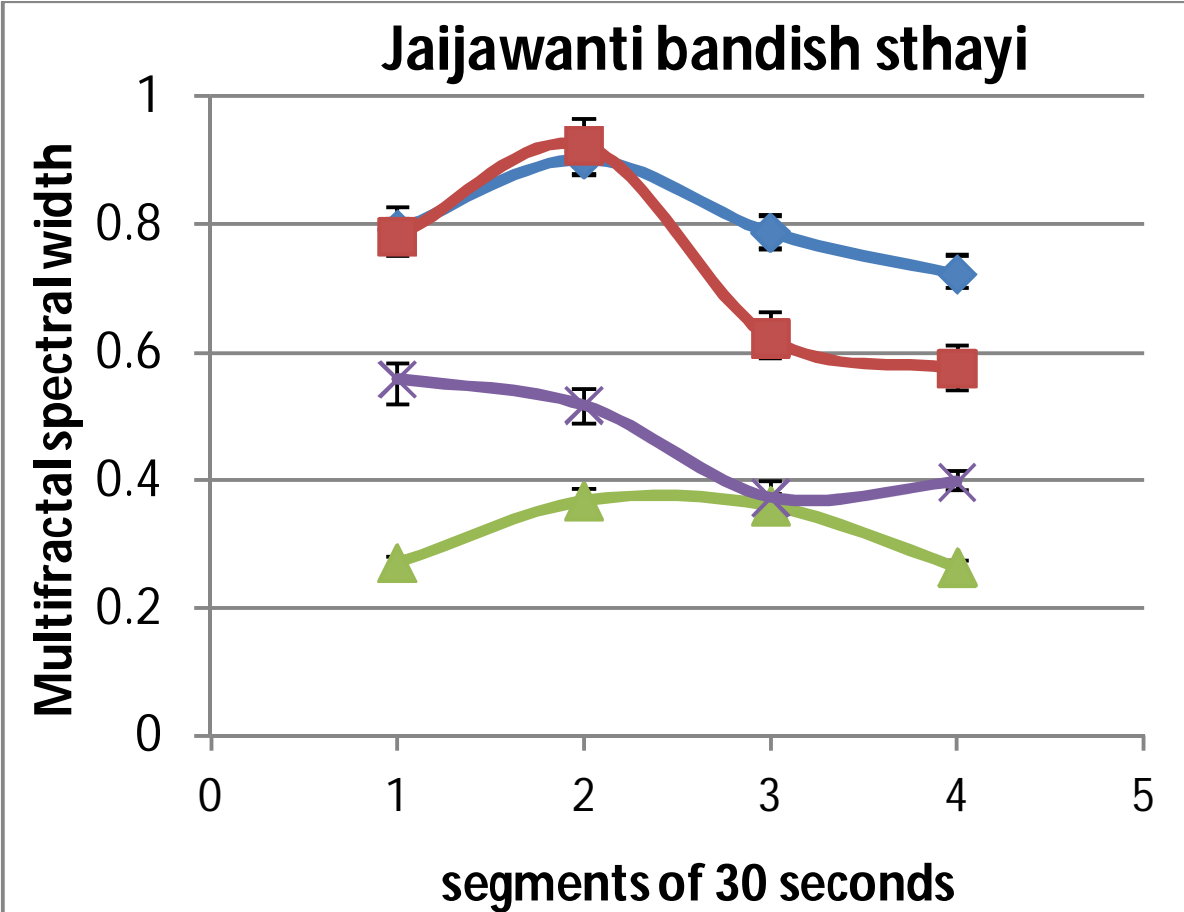

Fig. 2d

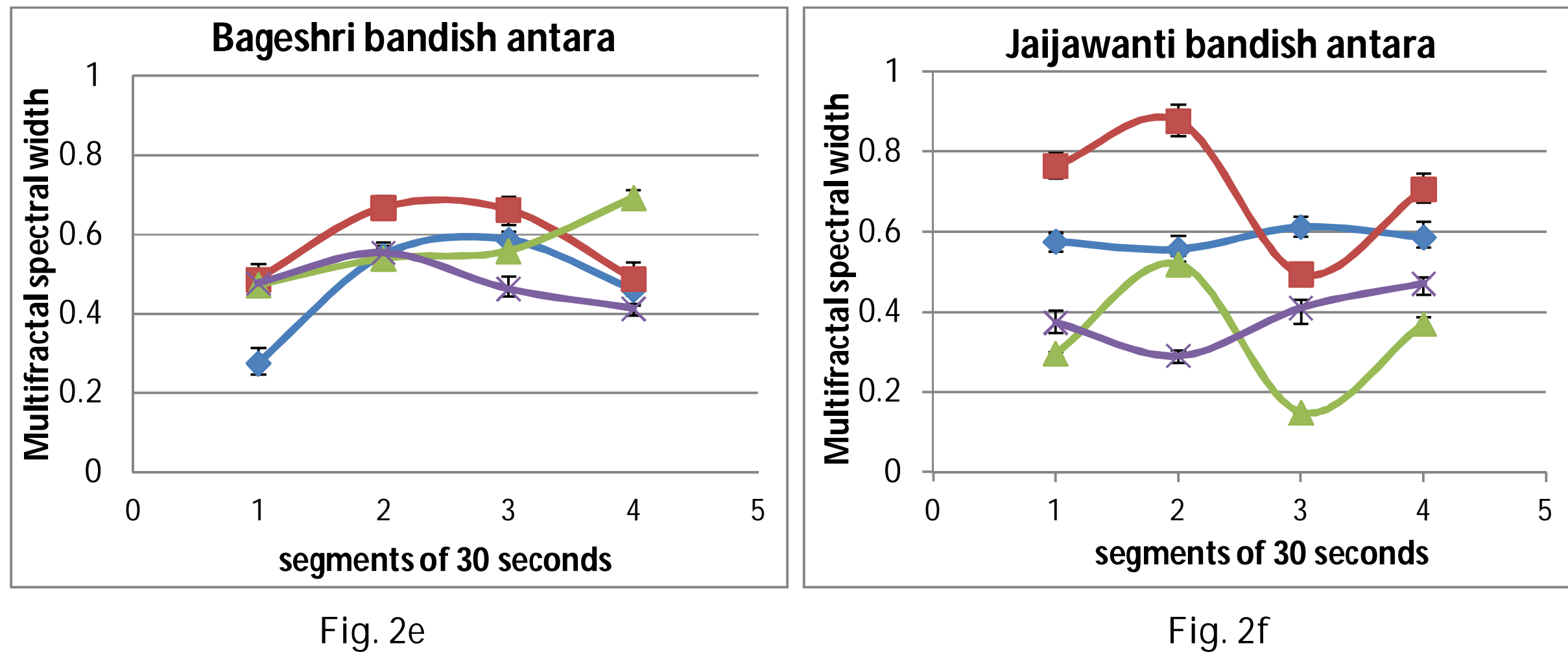

Fig. 2e Fig. 2f

**Fig. 2.** Variation of multifractal spectral widths in the (a, b) *Alap* parts; (c, d) *Bandish shtayi* parts and (e, f) *Bandish antara* parts of *Raga Bageshri* and *Raga Jaijawanti* respectively for the chosen four artists representing the four generations of artists from *Patiyala Gharana*.

The following observations can be drawn from a careful study of **Figures 2(a-f)**:

1. Comparing **Fig. 2a, Fig. 2c, Fig. 2e** with **Fig. 2b, Fig. 2d, Fig. 2f**, it can be easily observed that spectral width variation among the 4 artists of the same *Gharana* is higher in case of *Raga Jaijawanti* than *Raga Bageshri*. This complexity variation in case of *Raga Jaijawanti* is more prominent during *bandish sthayi* and *antara* parts compared to the *alap* part. These observations may be interpreted as following: In case of *Raga Bageshri* the singing style of the older generation artist is maintained more strictly by his successors than in case of *Raga Jaijawanti* where the new generation artists incorporated their own styles more frequently.
2. The complexity variation among the 4 artists is least in case of *Bageshri antara* (**Fig. 2e**) while largest in case of *Jaijawanti antara* (**Fig. 2f**).
3. In general for both *Raga*s, Artist 2 features greater average multifractal spectral width than other three artists both in *alap* part as well as *bandish* part.
4. Analysing the renderings of *Raga Bageshri* sung by the four artists, we can observe prominent similarities between Artist 1 and Artist 2 in the complexity variation pattern among the 4 parts of the *alap* section, but the artists of the newer generations (Artist 3 and Artist 4) differed from them following their own ways. In the *bandish sthayi* part, significant similarity was observed between Artist 2 and Artist 4, whereas Artist 1 and Artist 3 differed from them in a similar manner. In the *antara* part of the *Bageshri bandish*, a great degree of similarity was observed between the singing styles of Artist 1, Artist 2 and Artist 4, but Artist 3 slightly differed from all three of them. Though in *Raga Bagesgri*, the variation in spectral width among the four generations of artists is not very pronounced, but both in the *alap* section and the entire *bandish* section, Artist 1, Artist 3 and Artist 4 (i.e., the artists of first, third and fourth generation respectively) feature lower absolute values of multifractal spectral width compared to Artist 2 who is representing the second generation of artists from the *Patiyala Gharana*.
5. In case of *Raga Jaijawanti*, striking similarity is observed between Artist 1 and Artist 3 in the complexity variation pattern among the 4 parts of the *alap* as well as the *bandish sthayi* part whereas in *antara* part of the *bandish* Artist 3 resemble more with Artist 2 and Artist 4 resemble more with Artist 1, though the absolute values of the multifractal spectral width are much lesser for Artist 3 and Artist 4 compared to Artist 1 and Artist 2 in both *alap* and *sthayi* and *antara* parts of the *bandish*. So, it is evident that while singing *Raga Jaijawanti* the artists of the younger generations are trying to incorporate the singing style of the artists from the older generations as per their choice.

6. In *alap* part of both the *Raga*s Artist 4 significantly differs from other three artists, but in the *Bandish* (both in the *sthayi* and *antara* parts) he shows resemblance with other artists.
7. In most of the time segments we get varying complexity which has a clear tendency to increase from Artist 1 to Artist 2 but mostly decrease in case of the contemporary artists (Artist 3 and Artist 4 in our case).

Thus, the multifractal analysis of music signals can be efficiently used in analyzing the singing styles of different artists while performing any *Raga*.

## 5. CONCLUSION

Summarizing all the observations obtained from the multifractal detrended fluctuation analysis of the acoustical waveforms of the renderings in *Raga Bageshri* and *Raga Jaijawanti* by four artists of four consecutive generations belonging to the *Patiyala Gharana* of Hindustani classical music we can conclude that Artist 2 (who represents the second generation among the chosen four vocalists) seems to be completely different from the other three artists in *bandish* as well as *alap* parts in terms of complexity of the signal. Artist 3 and Artist 4, who represent the third and fourth generations of artists from *Patiyala Gharana* respectively, resemble Artist 1 of the oldest or first generation in specific time segments. From this trend we may predict that the contemporary artists Artist 3 and Artist 4 are trying to incorporate the style of Artist 1 in their singing sometimes. The variation in multifractal spectral width in different parts of the *alap* or *bandish* section of any *Raga* rendering reveals that in most of the time segments we get varying complexity which has a clear tendency to increase from Artist 1 to Artist 2 but mixed response or more commonly a decrement in the spectral width values in case of the contemporary artists. Analysis by this process using nonlinear chaos based multifractal techniques on the acoustical waveforms of different *Raga* renderings therefore yielded comparison of the singing styles among four vocalists of consecutive generations from a particular *gharana*, which serves our interest. Among all the chosen four artists, Artist 2 is both a direct disciple as well as a close blood relation family member of Artist 1. Artist 3 had the opportunity to learn *Raga* music directly from both Artist 1 and Artist 2 at different times. Artist 4 is a direct disciple of Artist 3. Now, on one hand, the similarities in the variation pattern of the multifractal spectral width (or simply the nonlinear acoustic complexity) in the *alap* and *bandish* sections of the renderings of *Raga Bageshri* and *Raga Jaijawanti* give us hint towards the preservation of some specific *Raga* rendering styles or the manifestation of *Guru-Shishya* tradition within a specific *Gharana* of Hindustani classical music. On the other hand, the dissimilarities in the singing pattern among these four artists indicate towards the improvisational tendencies of the individual artists which may also be affected by the ongoing globalization in every field of our daily lives.This is only a pilot study. Furthermore, our objective can be attained by the detailed analysis of the extracted notes used in the *Raga* and corresponding note to note transitions etc. Also, in future, the whole *alap* part as well as different other parts of several complete *Raga* performances should be analyzed and same study should be done for other *Gharanas* of Hindustani classical music to reach a more convincing conclusion.

## ACKNOWLEDGEMENT

Archi Banerjee acknowledges the Department of Science and Technology (DST), Govt. of India for providing the DST CSRI Post Doctoral Fellowship (DST/CSRI/PDF-34/2018) to pursue this research work. Shankha Sanyal acknowledges DST CSRI, Govt of India for providing the funds related to this Major Research Project (DST/CSRI/2018/78 (G)) and the Acoustical Society of America (ASA) for providing the International Students Grant.